# Scintillation and loss of signal lock from poleward moving auroral forms in the cusp ionosphere

K. Oksavik[1,2], C. van der Meeren[1], D. A. Lorentzen[1,2], L. J. Baddeley[1,2], and J. Moen[2,3]

[1]Birkeland Centre for Space Science, Department of Physics and Technology, University of Bergen, Bergen, Norway, [2]University Centre in Svalbard, Longyearbyen, Norway, [3]Department of Physics, University of Oslo, Oslo, Norway

**Abstract** We present two examples from the cusp ionosphere over Svalbard, where poleward moving auroral forms (PMAFs) are causing significant phase scintillation in signals from navigation satellites. The data were obtained using a combination of ground-based optical instruments and a newly installed multiconstellation navigation signal receiver at Longyearbyen. Both events affected signals from GPS and Global Navigation Satellite System (GLONASS). When one intense PMAF appeared, the signal from one GPS spacecraft also experienced a temporary loss of signal lock. Although several polar cap patches were also observed in the area as enhancements in total electron content, the most severe scintillation and loss of signal lock appear to be attributed to very intense PMAF activity. This shows that PMAFs are locations of strong ionospheric irregularities, which at times may cause more severe disturbances in the cusp ionosphere for navigation signals than polar cap patches.

## 1. Introduction

A fundamental characteristic of the dayside aurora is transients and poleward moving auroral forms (PMAFs). Early studies [e.g., *Feldstein and Starkov*, 1967; *Vorobjev et al.*, 1975] identified poleward moving events that detached from the dayside auroral oval and drifted into the polar cap. The motion of PMAFs is controlled by the polarity of the interplanetary magnetic field (IMF) $B_y$ component [*Sandholt et al.*, 1986, 1993; *Moen et al.*, 1999]. The PMAFs move northwest when $B_y$ is positive, and northeast when $B_y$ is negative [*Sandholt et al.*, 1998], and PMAFs are most frequent when $|B_y| > |B_z|$ [*Sandholt et al.*, 2004], corresponding to IMF clock angles between 45° and 135°.

PMAFs have often been interpreted as ionospheric signatures of flux transfer events (FTEs) [*Sandholt et al.*, 1990, 1993; *Denig et al.*, 1993; *Milan et al.*, 1999, 2000; *Thorolfsson et al.*, 2000], where dayside transients often show repetition rates [*Milan et al.*, 1999] comparable to FTEs at the magnetopause [*Russell and Elphic*, 1978, 1979; *Haerendel et al.*, 1978]. Transient reconnection is believed to be the primary transfer mechanism of flux from the solar wind to the magnetosphere [*Cowley and Lockwood*, 1992; *Lockwood et al.*, 1995]. A burst of reconnection propagates from the magnetopause to the ionosphere as an Alfvénic disturbance with an associated system of field-aligned Birkeland currents [*Glassmeier and Stellmacher*, 1996]. In the cusp ionosphere it sets up a mesoscale twin-cell flow pattern [*Southwood*, 1985, 1987]. The same process may also create isolated polar cap patches of increased electron density in the F region ionosphere [*Lockwood and Carlson*, 1992; *Carlson et al.*, 2002, 2004, 2006; *Lockwood et al.*, 2005a, 2005b; *Lorentzen et al.*, 2010], which are often closely associated with PMAFs and an optical flash at the foot of the newly opened flux [*Carlson et al.*, 2006].

A series of publications have used fast scan modes at the European Incoherent Scatter Svalbard Radar to investigate the time evolution of mesoscale flow channels in relation to PMAFs [*Carlson et al.*, 2004; *Oksavik et al.*, 2004, 2005, 2011; *Rinne et al.*, 2007, 2010; *Moen et al.*, 2008]. *Oksavik et al.* [2004, 2005] found that the clockwise vorticity on one side of a flow channel is consistent with an upward Birkeland current (i.e., intense PMAF aurora), while the counterclockwise vorticity on the other side of the flow channel is consistent with a downward Birkeland current (i.e., weak or no aurora). *Rinne et al.* [2007] identified a new type of flow channel (reversed flow events) that gives enhanced flow in the reverse direction of the large-scale background convection. *Moen et al.* [2008] point out that this phenomenon is related to a Birkeland Current Arc and provide two possible explanations: (1) coupling through a poorly conducting ionosphere of two MI current loops forced by independent voltage generators or (2) that the flow channel is driven by an inverted V.

The Super Dual Auroral Radar Network (SuperDARN) community has shown that the dayside aurora is often co-located with coherent HF radar echoes [*Rodger et al.*, 1995; *Yeoman et al.*, 1997; *Moen et al.*, 2001]. A wide range of poleward moving transients are observed: flow channel events [*Pinnock et al.*, 1993, 1995; *Chisham et al.*, 2000; *Neudegg et al.*, 1999, 2000], pulsed ionospheric flows (PIFs) [*Provan et al.*, 1998, 2002; *Provan and Yeoman*, 1999; *McWilliams et al.*, 2000] and poleward moving radar auroral forms [*Milan et al.*, 2000, 2002; *Wild et al.*, 2001; *Davies et al.*, 2002; *Rae et al.*, 2004]. SuperDARN obtains echoes from decametre-scale field-aligned plasma irregularities that track the background convection in the ionosphere [*Greenwald et al.*, 1995; *Chisham et al.*, 2007]. The close association between PMAFs and transient features in HF radar backscatter [*Milan et al.*, 1999] therefore suggests that PMAFs are associated with plasma irregularities, which may cause severe scintillation and disturbance of ground-to-satellite links and communication and navigation systems [*Buchau et al.*, 1985; *Basu et al.*, 1988, 1990, 1994, 1998].

Several studies have looked at scintillation of spacecraft signals in a statistical manner [*Kersley et al.*, 1995; *Spogli et al.*, 2009; *Li et al.*, 2010; *Alfonsi et al.*, 2011; *Tiwari et al.*, 2012]. Both 250 MHz satellite beacon scintillation measurements [*Aarons et al.*, 1981] and GPS scintillation measurements have found the highest occurrence of scintillation during the local winter months [*Li et al.*, 2010] or in the autumn-winter season [*Prikryl et al.*, 2011, 2015]. *Kersley et al.* [1995] pointed out that the occurrence of scintillation in winter often maximizes near magnetic noon and extends in a latitudinal belt into the afternoon/evening sector. Another study [*Prikryl et al.*, 2011] found maximum scintillation in the prenoon hours of the perturbed cusp ionosphere, in addition to nightside auroral arc brightening and substorms. Cusp region dynamics is proposed as a potentially strong source for phase scintillation and potential cycle slips [*Prikryl et al.*, 2010]. They defined a cycle slip as a jump in differential phase total electron content (TEC) of more than 1.5 TECU in 1 s (1 TECU corresponds to $10^{16}$ el/m$^2$). Scintillation and cycle slips have been found to peak when high-speed streams or interplanetary coronal mass ejections impact the Earth's magnetosphere [*Prikryl et al.*, 2014]. Intense auroral arcs have produced loss of signal lock during strong substorms in the nightside ionosphere [*Smith et al.*, 2008]; however, we have not yet found any reports in literature on loss of lock in the traditionally weaker dayside cusp aurora.

In a pioneering piece of work *Basu et al.* [1998] studied plasma structuring and scintillation over Svalbard for three active days in January 1997, but their work mostly focused on much larger spatial scales and the stable cusp aurora near magnetic noon. Later, *Milan et al.* [2005] found a close correspondence between the occurrence of amplitude scintillations of 250 MHz satellite beacon signals and SuperDARN backscatter power at 10 MHz. *Prikryl et al.* [2010, 2011] have also shown that GPS phase scintillation to coexist with SuperDARN backscatter. However, *Milan et al.* [2005] and *Prikryl et al.* [2010, 2011] have not related it to PMAF activity and moving auroral forms. Other case studies [*De Franceschi et al.*, 2008; *Coker et al.*, 2004; *Mitchell et al.*, 2005; *van der Meeren et al.*, 2014; *Jin et al.*, 2014; *Hosokawa et al.*, 2014] have mostly focused on nightside events in the context of magnetic storms or auroral substorms.

*Kinrade et al.* [2012] investigated ionospheric scintillation over Antarctica during a large geomagnetic storm following a coronal mass ejection. Significant phase scintillation was seen in the plasma depletion region both in the dayside noon sector and in the dayside cusp. Near 13:00 magnetic local time they observed ~30 s bursts of phase scintillation. Within each burst there were ~5–6 s pseudo-periodic oscillations. They suspected that this was due to "cusp precipitation of some kind," but no optical data or mentioning of PMAFs was presented.

*Kinrade et al.* [2013] compared auroral images with scintillation measurements from the South Pole Station by tracking up to 11 satellites simultaneously. At magnetic noon they found phase scintillation to be associated with 630.0 nm rather than 557.7 nm emissions. Summing over all local times (both day and night) they found that phase scintillations are generally more correlated with 557.7 nm than 630.0 nm emissions (correlation levels of up to 74% versus up to 63%). Their explanation was that 557.7 nm emissions have shorter lifetime and are generally more intense than 630.0 nm emission, which is usually not the case in the cusp region where 630.0 nm is expected to dominate [e.g., *Sandholt et al.*, 1986]. *Kinrade et al.* [2013] did not mention PMAFs or relate their dayside scintillation events to pulsed reconnection events at the magnetopause.

In the current paper we will follow up on the hypothesis of *Kinrade et al.* [2012] that the significant phase scintillation near magnetic noon is due to cusp precipitation of some kind. We will present data from a new receiver which is tracking data at Svalbard from GPS, Global Navigation Satellite System (GLONASS) and Galileo. In 2013, our receiver was on average tracking 20–24 satellites simultaneously. Using data from two intense PMAF events on 14 January 2013 we will document that PMAF activity is producing transient and highly localized areas of severe phase scintillation that move through the cusp region ionosphere and into the polar cap.

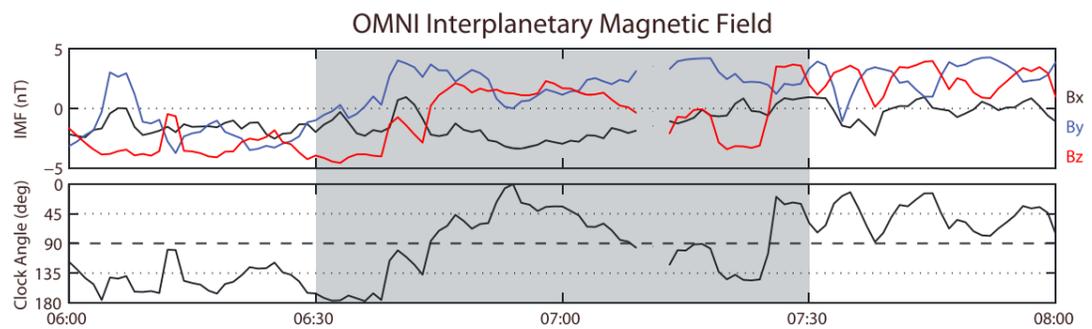

**Figure 1.** (top) Components of the interplanetary magnetic field (IMF): $B_x$ (black line), $B_y$ (blue line), $B_z$ (red line). (bottom) The IMF clock angle is also shown. Grey shading is used to indicate the time interval shown in Figure 2.

## 2. Instrumentation

For this study we use a newly installed NovAtel GPStation-6 receiver at the Kjell Henriksen Observatory (KHO) in Longyearbyen (78.1°N, 16.0°E). It is a multiconstellation and multifrequency receiver, which is currently tracking signals from GPS (L1/L2/L2C/L5), GLONASS (L1/L2), and Galileo (E1/E5a/E5b/Alt-BOC). The receiver outputs the phase scintillation $\sigma_\phi$ index [*Fremouw et al.*, 1978; *Rino*, 1979]. A sixth-order Butterworth high-pass filter with a cutoff frequency of 0.1 Hz is used to find the detrended raw carrier phase $\phi$, and the $\sigma_\phi$ index is computed over 60 s intervals [*van Dierendonck et al.*, 1993, 1996]:

$$\sigma_\phi^2 = \langle \phi^2 \rangle - \langle \phi \rangle^2$$

The phase scintillation index is generally influenced by the observation geometry, but *Forte and Radicella* [2004] have shown that geometrical factors are not important for scintillation indices at high latitudes for satellites flying in GPS-like orbits. The receiver also outputs the 60 s amplitude scintillation $S_4$ index, which is the standard deviation of the received power $I$ normalized by its mean value [*Briggs and Parkin*, 1963]:

$$S_4^2 = \frac{\langle I^2 \rangle - \langle I \rangle^2}{\langle I \rangle^2}$$

The receiver also provides the total electron content (TEC) and rate of TEC (ROT), both at 1 s and 60 s resolution. Raw data of the amplitude and phase are available at 50 Hz resolution.

For optical monitoring of the PMAF activity we use a meridian scanning photometer (MSP) and an all-sky imager (ASI) both located at KHO. The geographic location of KHO relative to the magnetic pole allows for detailed observations of the dayside aurora in the midst of winter. The MSP is operated by the University Centre in Svalbard, and it was recording auroral emissions at 557.7 and 630.0 nm. The ASI is operated by the University of Oslo, and it was recording auroral emissions at 630.0 nm. We will also use solar wind data from the NASA OMNIWeb service, which provides data that are already time-shifted to the bow shock.

## 3. Data Presentation

Around 07:00 UT on 14 January 2013 the solar wind speed was 520–560 km/s and the solar wind density was 2–3 cm$^{-3}$ (data not shown). The Earth's geomagnetic field was weakly disturbed ($Kp = 3$). Figure 1 presents interplanetary magnetic field (IMF) data that have been extracted from the NASA/ Goddard Space Flight Center's OMNI data set. Figure 1 (top) shows the components $B_x$ (black line), $B_y$ (blue line), and $B_z$ (red line). Although there is a brief data gap around 07:10 UT, the figure shows that $B_x$ is negative and both $B_y$ and $B_z$ are weakly positive around 07:00 UT. Figure 1 (bottom) shows the IMF clock angle in the GSM y-z plane [see, e.g., *Oksavik et al.*, 2000]. It is defined as $\theta = \tan^{-1}(|B_y/B_z|)$ for $B_z > 0$ and $180° - \tan^{-1}(|B_y/B_z|)$ for $B_z < 0$. In the interval 06:45–07:10 UT the IMF clock angle is between 0° and 90°. The positive $B_y$ is favorable for PMAF activity over Svalbard with motion from southeast to northwest [*Sandholt et al.*, 1998, 2004].

Figure 2 gives an overview of optical and scintillation data from the Kjell Henriksen Observatory (KHO) in Longyearbyen. Figures 2a and 2b show the intensity of the meridian scanning photometer (MSP) 630.0 nm and 557.7 nm channels, respectively. The MSP scans the sky from north (0° elevation) to south (180° elevation) with a time resolution of 16 s. The background has been subtracted, and the color scale gives the intensity in Rayleigh ($R$). Up until 06:55 UT there was a bright arc in the southern part of the sky (around 120–150° elevation), which is characteristic for the strongly southward IMF dayside cusp aurora [*Sandholt et al.*, 1998, 2004]. At 06:55–07:00 UT and 07:05–07:15 UT there were two PMAFs that formed and drifted into the polar cap. These events are indicated with dashed lines and numbers 1 and 2 and occurred for positive IMF $B_y$, which is favorable for PMAF activity [*Sandholt et al.*, 1998, 2004].

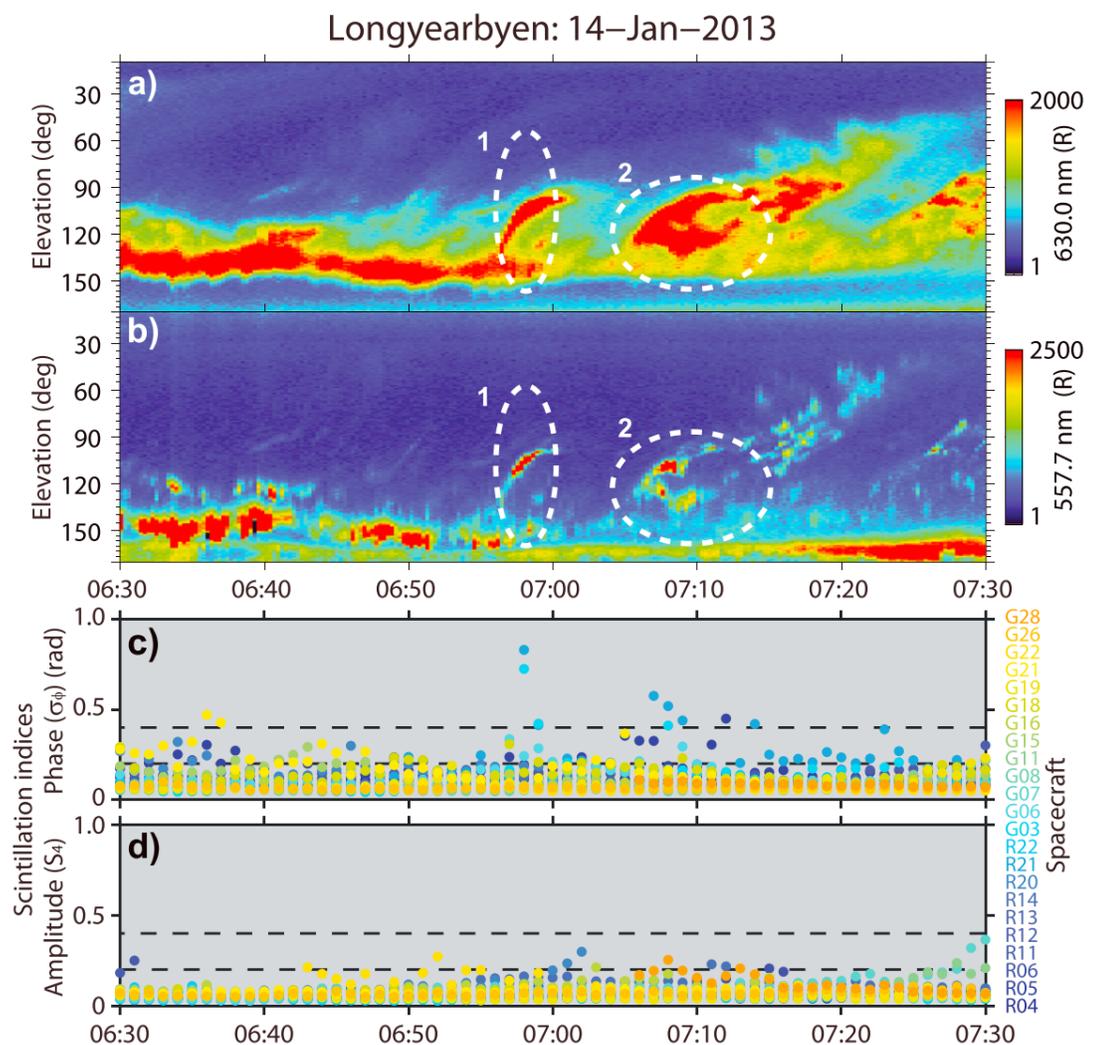

**Figure 2.** Optical and scintillation data from the Kjell Henriksen Observatory (KHO) in Longyearbyen: (a) the 630.0 nm meridian scanning photometer (MSP) intensity, (b) the 557.7 nm MSP intensity, (c) the 60 s phase scintillation $\sigma_\varphi$ index, and (d) 60 s amplitude scintillation $S_4$ index. Colors are used in Figures 2c and 2d to separate the different spacecraft. The dashed lines in Figures 2a and 2b indicate the PMAF events 1 and 2. See text for more details.

Figures 2c and 2d show the corresponding scintillation data from our new multiconstellation navigation signal receiver at KHO in Longyearbyen, and colors are used to separate the different spacecraft. Figure 2c shows the 60 s phase scintillation $\sigma_\phi$ index. Phase scintillations are caused by irregularities of scale size from hundreds of meters to several kilometers [*Kintner et al.*, 2007]. In Figure 2c we notice that both PMAF events coincide in time with two intervals of enhanced phase scintillation ($\sigma_\phi$ = 0.2 to 0.8 radians). The first event had enhanced phase scintillation lasting for 3 min (06:57–06:59 UT). The second event had enhanced phase scintillation lasting for 5 min (07:05–07:09 UT) and an additional 3 min (07:11–07:14 UT). Figure 2d shows the 60 s amplitude scintillation $S_4$ index. In the weak scattering regime [*Rino*, 1979] of these observations, amplitude scintillations at L band are caused by irregularities of scale size from tens of meters to hundreds of meters, which is at and below the Fresnel radius [*Kintner et al.*, 2007]. In Figure 2d we notice that the amplitude scintillation $S_4$ index is generally less than 0.2, which is typical for high latitudes where amplitude scintillations are weak and phase scintillations dominate [*Kintner et al.*, 2007].

Figure 3 presents six images (06:55–07:00 UT) from the University of Oslo all-sky imager at KHO in Longyearbyen. In Figure 3a the image has been projected onto a magnetic grid, assuming auroral emissions at 250 km altitude, which is a typical altitude for 630.0 nm emissions in the cusp ionosphere [*Lockwood et al.*, 1993; *Johnsen et al.*, 2012], while the other panels show unprojected all-sky images (north up, east right). The color scale indicates the auroral intensity at 630.0 nm (lower color bar). Overlaid onto each panel we also show scintillation data, where shapes indicate the type of constellation: GLONASS (diamonds) and GPS (circles). The marker color is the phase scintillation index (upper color bar). Both the auroral and phase scintillation activity is low in the first two panels (06:55–06:56 UT). At 06:57 UT a bright PMAF appeared, and the phase scintillation immediately began to rise in its vicinity (GLONASS 21 and GPS 03). At 06:58 UT the PMAF had intensified further and began drifting northwest. GPS 03 and GLONASS 21 showed strong phase scintillation right in the middle of the PMAF. At 07:00 UT the PMAF had left the area covered by GPS and GLONASS, and the phase scintillation returned back to low levels. It should also be pointed out that throughout the entire time interval 06:55–07:00 UT the phase scintillation only changed in the vicinity of the PMAF. The phase scintillation was weak in the rest of the field of view. Consequently, the area of strong phase scintillation and PMAF activity appears to be related.

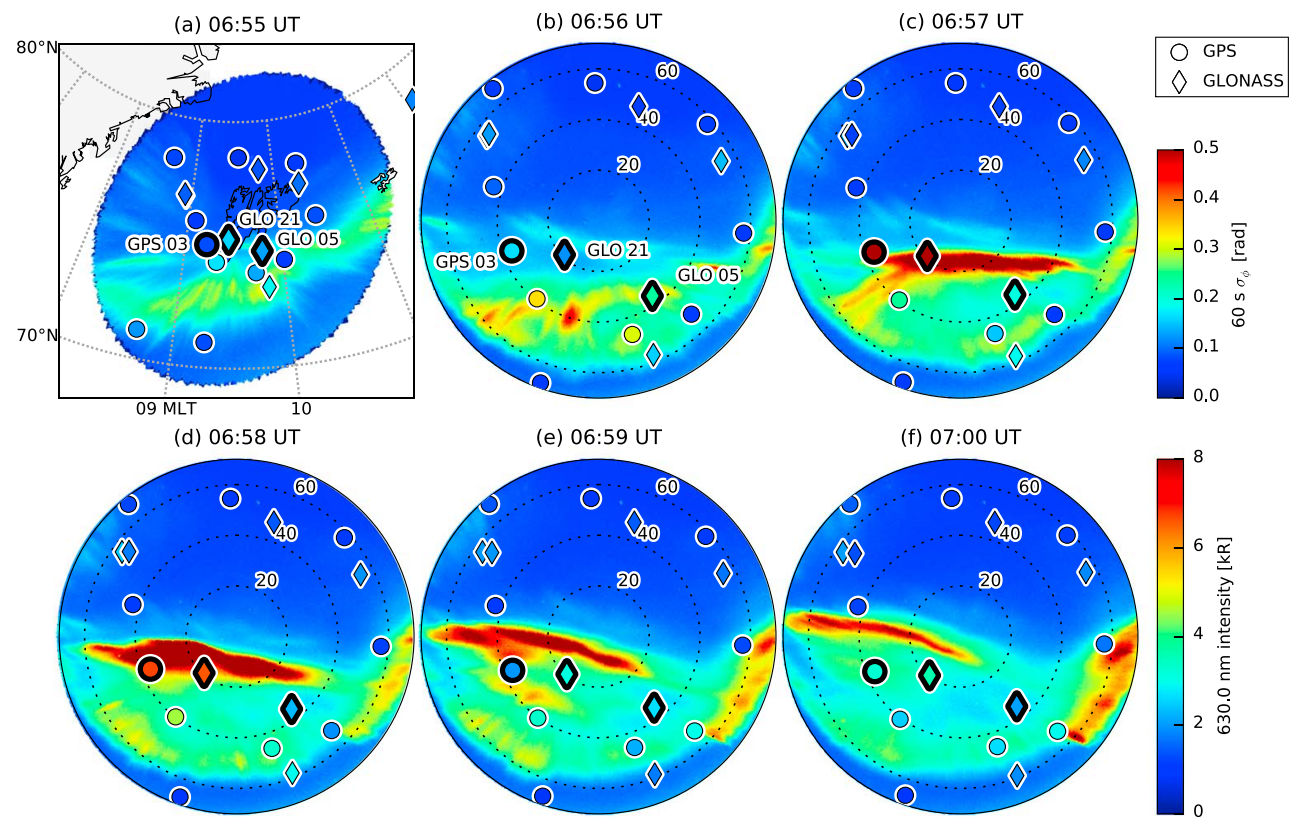

**Figure 3.** Six 630.0 nm all-sky images from the Kjell Henriksen Observatory (KHO) in Longyearbyen on 14 January 2013 with ionospheric piercing points (IPPs) overlaid for all available GPS and GLONASS spacecraft. In the first panel (a) the image is projected to a magnetic grid, while the other panels show unprojected all-sky images (magnetic north is up, east is right). The marker color is the phase scintillation index (upper color bar). Intense phase scintillation occurred when the GPS 03 and GLONASS 21 signals intersected a bright PMAF between 06:57 and 06:58 UT.

Figure 4 presents another example in the same format. The first image frame (07:03 UT) shows a bright PMAF east of Svalbard. At 07:04 UT the PMAF begins to intersect GLONASS 05, which shows elevated phase scintillation. The PMAF moves northwest. At 07:06 UT GLONASS 05 continues to show enhanced phase scintillation, while GLONASS 21 starts to show strong phase scintillation. At 07:07 UT the PMAF has intensified and moved further northwest, and both GPS 03 and GLONASS 21 show strong phase scintillation. At 07:08 UT the PMAF is located more to the northwest. GPS 03 is no longer affected, while GLONASS 21 continues to show strong phase scintillation. Consequently, the area of strong phase scintillation again overlaps with the PMAF.

Figures 5, 6, and 7 provide a closer look at the phase scintillation and TEC variations in relation to line-of-sight auroral emissions for the three spacecraft that were highlighted in Figures 3 and 4 (GLONASS 05, GLONASS 21, and GPS 03). The grey shading indicates the intervals covered by Figures 3 and 4 for easier comparison.

Figures 5a, 6a, and 7a present the line-of-sight auroral intensity at 557.7 and 630.0 nm in an area that is $7 \times 7$ pixels wide and centered at the elevation and azimuth of each spacecraft. The width of each line represents the range of observed auroral intensities in the $7 \times 7$ pixel area (from minimum to maximum intensity). For both events the GLONASS 05 (Figure 5a) signal was experiencing significantly lower auroral intensities than GLONASS 21 (Figure 6a) and GPS 03 (Figure 7a), which is consistent with GLONASS 05 being slightly equatorward of the two PMAF events. For GLONASS 05 (Figure 5a) the intensity ratio 630.0 nm versus 557.7 nm stayed around 2:1, which is typical for the dayside aurora, where 630.0 nm usually dominates [e.g., *Sandholt et al.*, 1986]. For GLONASS 21 (Figure 6a) and GPS 03 (Figure 7a) the intensity ratio stayed

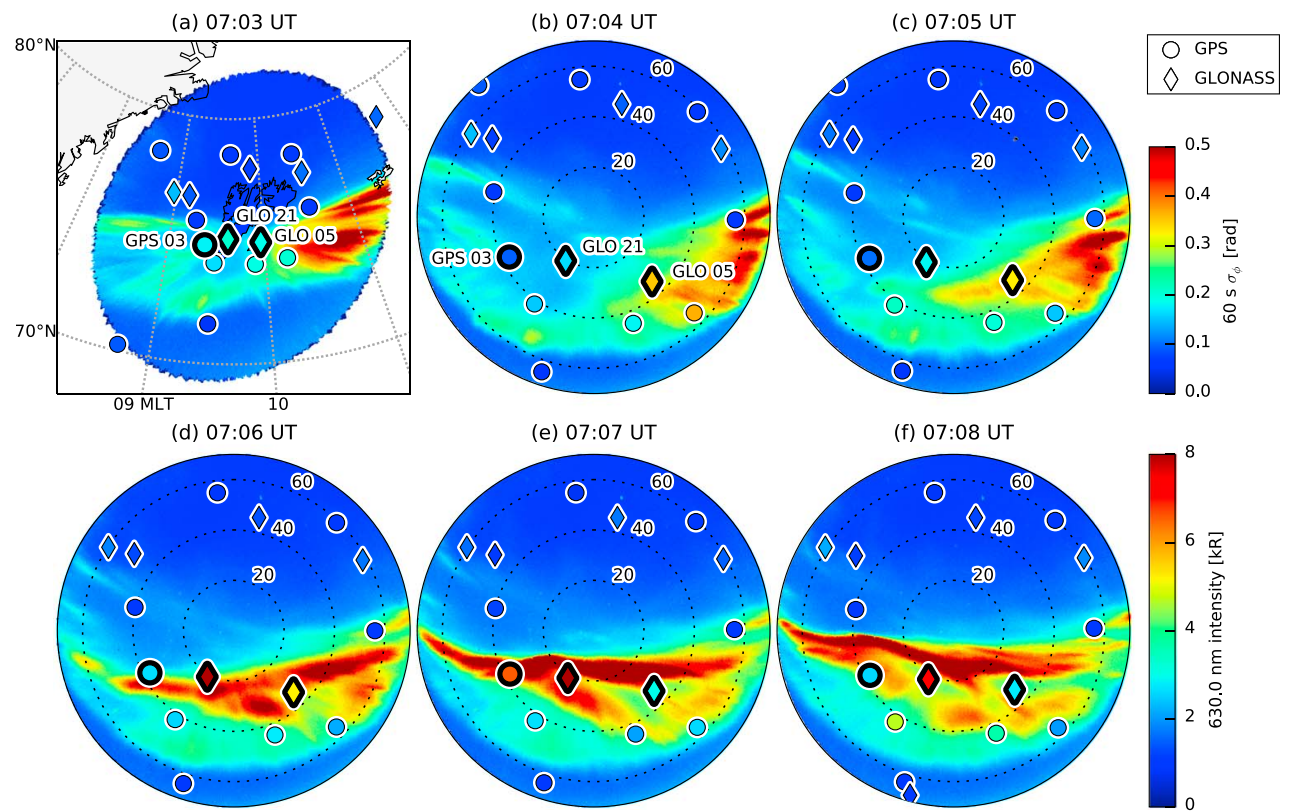

**Figure 4.** Same as Figure 3 but for 07:03–07:08 UT on 14 January 2013. Intense phase scintillation occurred when the GPS 03 and GLONASS 21 signals intersected a bright PMAF between 07:06 and 07:08 UT.

around 2:1 before/after the two PMAF events. In the middle of each PMAF event the auroral intensity spiked for both 630.0 and 557.7 nm, and the intensity ratio approached 1:1. For GPS 03 (Figure 7a) the 557.7 nm intensity also exceeded the 630.0 nm intensity for a few seconds around 06:57:30 UT.

Figures 5b and 5c, 6b and 6c, and 7b and 7c present the TEC and the absolute value of the rate of TEC (ROT), respectively. All spacecraft were at high elevations, so there is generally little difference between the slant TEC (solid lines) and the vertical TEC (dashed lines). All TEC data have been calibrated (corrected for bias). In the TEC data (Figures 5b, 6b, and 7b) there were several polar cap patches (indicated with the letter P), which can be identified as transient TEC enhancements lasting just a few minutes. For GLONASS 05 (Figures 5a and 5b) the PMAFs at 06:56 and 07:06 UT were co-located with local TEC minima, that were followed by polar cap patches (i.e., the PMAFs were on the poleward side of the polar cap patch, given that the drift speed was in the poleward direction in Figures 3 and 4). For GLONASS 21 (Figures 6a and 6b) the first PMAF at 06:57 UT was co-located with a TEC minimum, while the second PMAF at 07:07 UT was inside a polar cap patch.

For GPS 03 (Figures 7a and 7b) the PMAF at 07:07 UT was co-located with a TEC minimum, while the PMAF at 06:57 UT coincided with a dramatic TEC enhancement. From 06:57:13 to 06:57:20 UT the TEC jumped from 2.8 to 18.9 TECU, corresponding to a gradient of 2.3 TECU/s, which according to *Prikryl et al.* [2010] would be classified as a cycle slip (change of more than 1.5 TECU/s). A careful examination of the raw data also reveals that the receiver lost lock for the L2Y signal (center frequency at 1227.60 MHz) between 06:56:44 and 06:57:36 UT (indicated with cyan shading in Figure 7b). There was no loss of lock for L1CA, which corresponds to a higher center frequency (1575.42 MHz). Unfortunately, the receiver was only recording 50 Hz data of L1CA at the time, but 1 s resolution TEC data, which were recorded in real-time and were based on L1CA and L2Y signals, reveal two 5 and 3 s data gaps (06:57:13 to 06:57:18 UT and 06:57:33 to 06:57:36 UT). It suggests that the L2Y loss of lock occurred at those particular times. It should be noticed that 06:57:13 to 06:57:18 UT coincides both with the extremely bright 557.7 and 630.0 nm PMAF (Figure 7a) and the steep TEC gradient (Figure 7b) that *Prikryl et al.* [2010] would classify as a cycle slip.

For GLONASS 05 the ROT (Figure 5c) was generally lower than for the other two spacecraft (see Figures 6c and 7c). GLONASS 05 observed the highest ROT of around 0.25 TECU/s in the vicinity of local TEC minima (e.g., 06:56, 06:59, 07:06, and 07:11 UT), often in connection with a positive TEC gradient. Given the poleward drift seen in Figures 3 and 4, it suggests the highest ROT was for the most part detected at the poleward side (leading edge) of polar cap patches. Only two of the ROT enhancements appear to be related to small enhancements of auroral emissions (06:56 and 07:06 UT). Both GLONASS 21 and GPS 03 had their highest ROT in connection with the two PMAFs at 06:57 and 07:07 UT. For GLONASS 21 the ROT exceeded 0.5 TECU/s (Figure 6c), and for GPS 03 the ROT exceeded 0.2–0.3 TECU/s (Figure 7c).

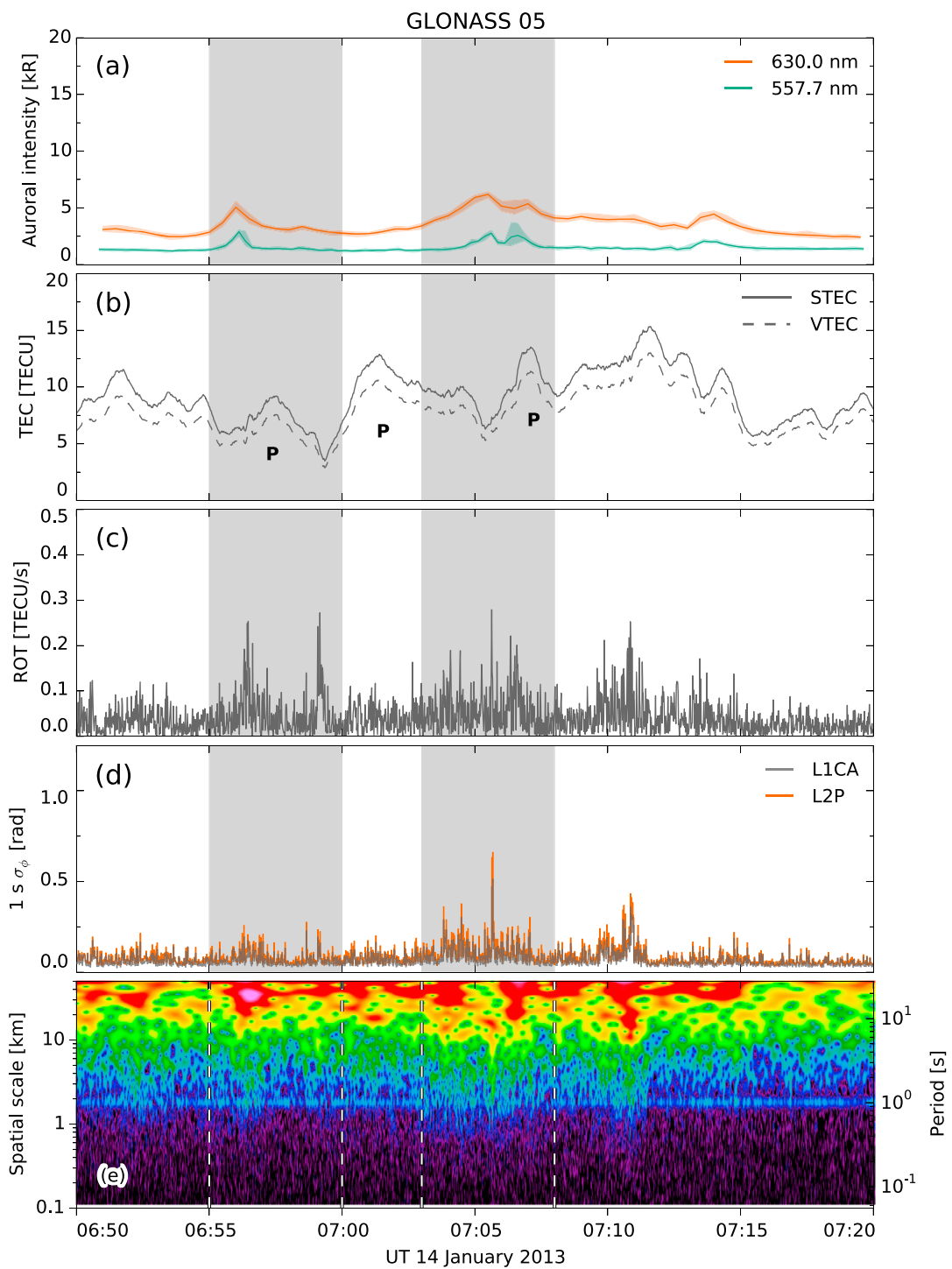

**Figure 5.** GLONASS 05 data on 14 January 2013. (a) The 557.7 and 630.0 nm auroral intensity in the vicinity of the satellite IPPs. (b): Slant and vertical total electron content (TEC) in units of TECU (1 TECU = $10^{16}$ el m$^{-2}$). Three polar cap patches are indicated using the letter P. (c) Rate of TEC (ROT). (d) The phase scintillation index computed from raw 50 Hz phase data over periods of 1 s for the signals L1CA and L2P. (e): Wavelet power spectra of raw 50 Hz phase data versus the corresponding period (right axis) and spatial scale (left axis). Grey shading indicates the duration of Figures 3 and 4. The fluctuations are generally lower than for the other two spacecraft (see Figures 6 and 7).

Figures 5d, 6d, and 7d present a high-resolution $\sigma_\phi$ scintillation index that we have calculated to provide a more detailed view of the phase scintillation during the events. The raw 50 Hz phase data were detrended using a sixth-order Butterworth high-pass filter with a cutoff frequency of 0.1 Hz, and the $\sigma_\phi$ index was computed over 1 s intervals. We have carefully examined that this high-resolution $\sigma_\phi$ index matches the overall features of the lower-resolution 60 s data, which is output by the receiver in real time. For GLONASS 05 (Figure 5d) and GLONASS 21 (Figure 6d) the 1 s $\sigma_\phi$ index was computed for both L1CA and L2P. For GPS 03 (Figure 7d) we could only calculate it for L1CA, because L2Y was not recorded.

For GLONASS 05 (Figure 5d) the phase scintillation was relatively low, except for a brief enhancement of $\sigma_\phi > 0.5$ rad around 07:06 UT in connection with the second PMAF event. The otherwise weak scintillation is consistent with GLONASS 05 being slightly equatorward of the PMAF activity. For GLONASS 21 (Figure 6d) and GPS 05 (Figure 7d) the phase scintillation peaked in connection with the two PMAF events at 06:57 and 07:07 UT, reaching $\sigma_\phi \sim 1.0$ rad for GLONASS 21, and $\sigma_\phi \sim 0.5$ rad for GPS 03. Both are indicative of severe phase scintillation in connection with the bright PMAFs. We should also point out that for GLONASS 05 and 21 (Figures 5d and 6d) the phase scintillation was generally higher at L2P (center frequency at 1242–1248 MHz) than at L1CA (center frequency at 1598–1605 MHz), which is indicative of irregularities being stronger at longer spatial scales (lower frequency corresponds to longer wavelength). We will investigate this next.

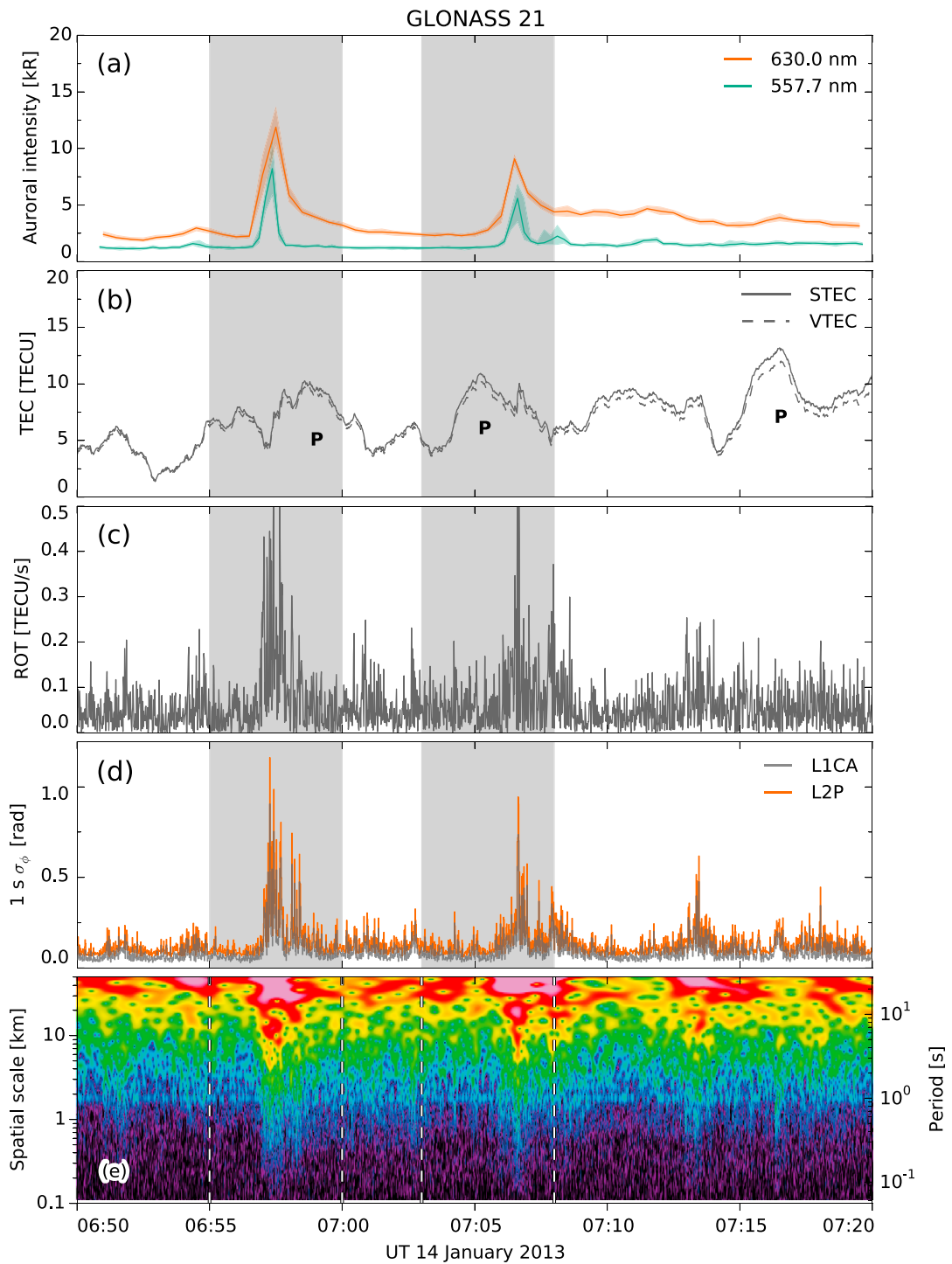

**Figure 6.** Same as Figure 5 but for GLONASS 21. Notice the high auroral intensities, enhanced fluctuations of phase and TEC, and the spectrum extending down to spatial scales of a few hundreds of meters during the two intervals highlighted with grey shading.

Figures 5e, 6e, and 7e present spectrograms of the raw phase in order to obtain more detailed information on the phase variations in relation to spatial scale size, using a similar approach to *van der Meeren et al.* [2014]. The axes on the right side show the period of the phase variations. The spectrograms were made using a wavelet analysis, based on software provided by *Torrence and Compo* [1998]. The Morlet wavelet was chosen as the mother wavelet. This method has previously been used by other GNSS studies [e.g., *Mushini et al.*, 2012]. Some key advantages of the wavelet technique are that (1) no detrending of the GNSS data is required

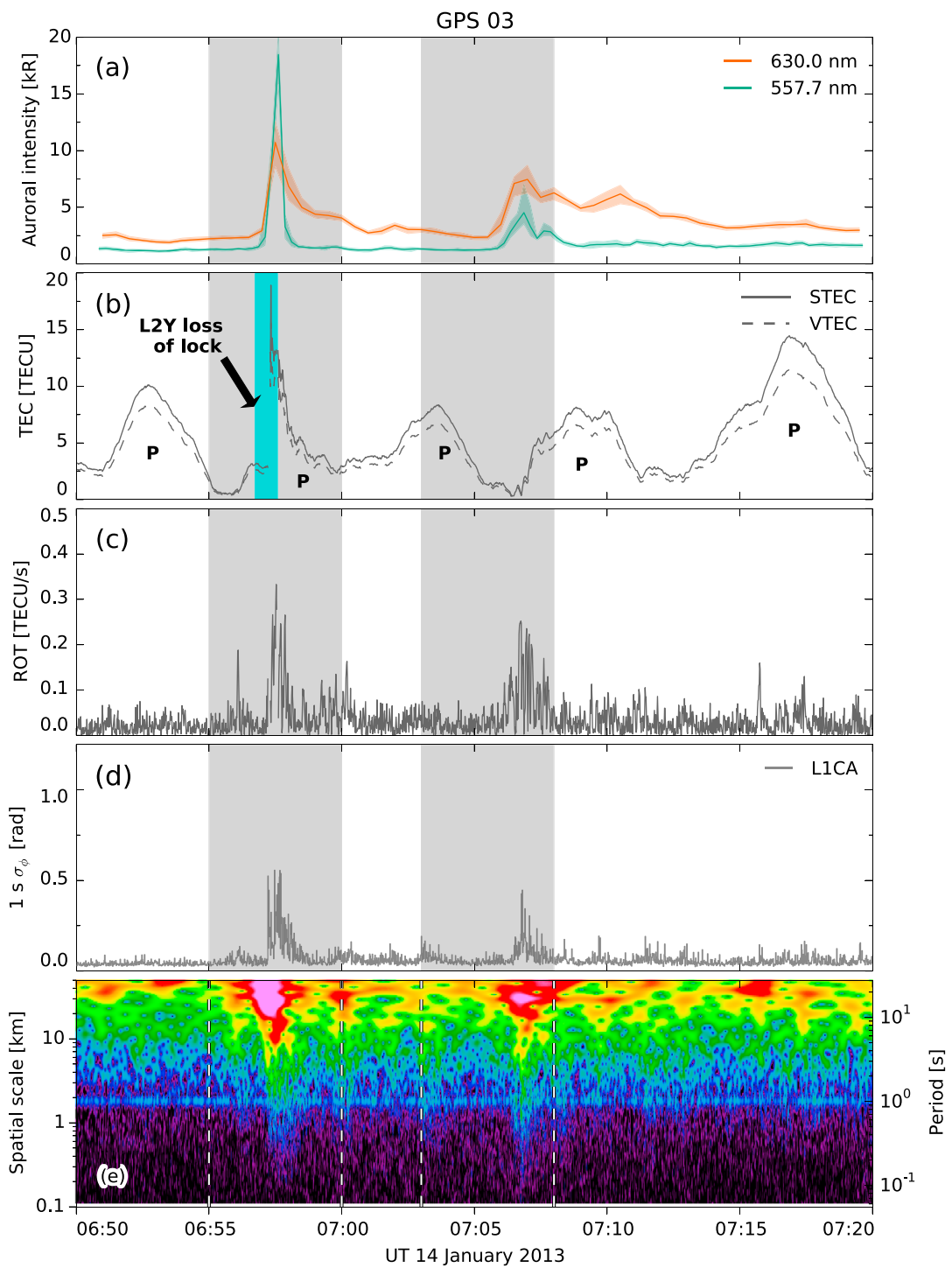

**Figure 7.** Same as Figure 5 but for GPS 03. The phase scintillation index in Figure 7d is computed using the L1CA signal. Notice the high auroral intensities, enhanced fluctuations of phase and TEC, and the spectrum extending down to spatial scales of a few hundreds of meters during the two intervals highlighted with grey shading. A loss of lock was observed for the first PMAF event (cyan shading in Figure 7b).

to produce a wavelet spectrogram and (2) wavelet spectrograms provide better resolution at smaller scales. The spectrograms have been carefully examined and compared to Fourier spectrograms of detrended data to verify that the two techniques give similar results. For further details on the wavelet technique we refer to *Torrence and Compo* [1998] and *Mushini et al.* [2012].

The wavelet spectrograms were converted to spatial scale (see left axes), using information on the drift speed of the PMAFs. From an analysis of all-sky images we found that two PMAFs had drift speeds of around 1600 and 2300 m/s, respectively. At the same time the ionospheric piercing points of the satellites moved at 40–50 m/s, which is insignificant in this regard. An average value of 1800 m/s was chosen for the conversion of temporal features to a spatial scale. Note that the spatial scale is only used in an order-of-magnitude sense (i.e., an adjustment to the assumed drift velocity by a factor 2 will only cause a corresponding linear adjustment to the spatial scale in the spectrogram, which will not change the order of magnitude).

For GLONASS 05 (Figure 5e) the strongest spectral power enhancements (red and yellow colors) were seen at spatial scales greater than 10 km, and some spectral power enhancements (green and bright blue colors) extended all the way down to ~2 km spatial scale. This picture was quite uniform throughout the whole time interval, which is consistent with GLONASS 05 being less affected by PMAF activity. For the most intense PMAF activity around 06:57 and 07:07 UT, GLONASS 21 (Figure 6e) and GPS 03 (Figure 7e) showed strong spectral power enhancements (red and yellow colors) at spatial scales greater than 4–5 km, and some spectral power enhancements (green and bright blue colors) extended all the way down to just a few hundreds of meters. Consequently, the finest structuring appears to be highly localized and attributed to PMAFs. It suggests that PMAFs are locations of stronger irregularities than the surrounding cusp ionosphere, which also included several polar cap patches (Figures 5b, 6b, and 7b), in particular at spatial scales of a few hundreds of meters to a few kilometers.

## 4. Discussion

Our results show that both GPS and GLONASS signals were affected by PMAF activity. The enhanced scintillation in the cusp ionosphere was observed to be highly localized and highly transient in nature. This is contrary to the results of *Prikryl et al.* [2011], who reported that scintillation often covers a large geographic area of the cusp/cleft region and with duration of several hours. While most of the field of view was unaffected, the phase scintillation was enhanced in the close vicinity of the two PMAF events. It therefore appears that the cusp precipitation of some kind, which *Kinrade et al.* [2012] were referring to as the cause of significant phase scintillation near magnetic noon, must be the well-known phenomenon in the dayside aurora called PMAFs. The area of enhanced phase scintillation drifted poleward into the polar cap together with the PMAF. The motion of the two PMAF events is consistent with reports on PMAFs in literature [e.g., *Sandholt et al.*, 1998].

The introduction section pointed out that PMAFs are often closely associated with polar cap patches [*Carlson et al.*, 2006], which also are known to follow the antisunward convective flow across the polar cap. Polar cap patches are known to have densities 2–20 times higher than the surrounding background electron density [*Buchau et al.*, 1983; *Weber et al.*, 1984; *Crowley et al.*, 2000]. They form near the dayside polar cap boundary and drift across the polar cap to the nightside [*Lorentzen et al.*, 2004; *Oksavik et al.*, 2010; *Moen et al.*, 2013; *Nishimura et al.*, 2014; *Zhang et al.*, 2015]. Polar cap patches are known to cause field-aligned plasma irregularities [e.g., *Hosokawa et al.*, 2009], and several mechanisms have been proposed in literature.

One example is plasma density gradients that by themselves can grow unstable due to the gradient drift (GD) plasma instability [*Ossakow and Chaturvedi*, 1979; *Keskinen and Ossakow*, 1983; *Tsunoda*, 1988; *Basu et al.*, 1988, 1990, 1994, 1998; *Gondarenko and Guzdar*, 2004], which is often regarded as the dominant mode for production of electron density irregularities in the *F* region cusp. The GD plasma instability mechanism occurs for plasma drift, of the correct sign, across a steep plasma density gradient perpendicular to the Earth's magnetic field at high latitudes [*Keskinen and Ossakow*, 1983]. Plasma drift in the opposite direction will set up polarization fields that stabilize the plasma against formation of irregularities. Recent sounding rocket data have shown that decametre-scale irregularities are located on kilometre-scale electron density gradients in the cusp ionosphere produced by electron precipitation, with estimated growth times of 10–50 s for the GD process [*Moen et al.*, 2012].

Another example is shears and vorticity in the plasma flow that are associated with PMAFs [*Oksavik et al.*, 2004, 2005, 2011; *Rinne et al.*, 2007, 2010; *Moen et al.*, 2008]. Flow shears are known to trigger the Kelvin-Helmholtz (KH) plasma instability [*Basu et al.*, 1988, 1990; *Keskinen et al.*, 1988]. The KH theory has been further developed by *Keskinen et al.* [1988], who also included a refinement of ionosphere-magnetosphere electrical coupling. Using SuperDARN observations of flow shears, *Oksavik et al.* [2011] found KH irregularity growth times of 1–5 min. Several of their events were associated with wide Doppler spectra and enhanced backscatter power, consistent with the growth of plasma irregularities.

There have also been attempts to merge the GD and KH plasma instability mechanisms in a two-step process. *Carlson et al.* [2007, 2008] proposed initial structuring driven by the KH instabilities, followed by additional structuring down to much finer scales driven by the GD instabilities. *Oksavik et al.* [2012] studied sounding rocket data from the cusp ionosphere, revealing plasma irregularities extending from hundreds of meters down to a few tens of meters. However, the KH mechanism could not explain the finest-scale irregularities. *Oksavik et al.* [2012] noticed that the strongest plasma irregularities were observed 2 min after a significant enhancement in the aurora and proposed an alternative two-step process: (1) structured particle precipitation first generates weak "seed" irregularities and (2) the GD instability then breaks these seed irregularities down to smaller scales.

The cases we presented have several polar cap patches (indicated with the character P in Figures 5b, 6b, and 7b). However, it is surprising to notice that most patches did not produce enhanced scintillation. One would normally expect the plasma inside a polar cap patch to become fully structured soon after initiation and with irregularities extending throughout the whole patch [*Hosokawa et al.*, 2009]. Alternatively, one would at least expect irregularities at the steep gradients near the trailing edge of polar cap patches, which is unstable to irregularity growth via the GD mechanism [e.g., *Milan et al.*, 2002]. However, our observations do not support any of these two options. Both Figures 6 and 7 show the strongest phase scintillation to be co-located with extremely bright PMAFs, suggesting that structured particle precipitation is a very important source for plasma irregularities, at least at kilometer to hundred meter scale [*Oksavik et al.*, 2012]. PMAFs are also associated with severe flow shears [*Oksavik et al.*, 2004, 2005, 2011; *Rinne et al.*, 2007, 2010; *Moen et al.*, 2008], which may contribute to irregularities via the KH mechanism [*Oksavik et al.*, 2011].

The observed phase scintillations are believed to be due to irregularities of scale size from hundreds of meters to several kilometers [*Kintner et al.*, 2007]. Our observations in Figures 6e and 7e show that irregularities at these spatial scales are present when PMAFs intersect the signal path. Sounding rocket data [*Moen et al.*, 2012; *Oksavik et al.*, 2012] also document that irregularities can exist all the way down to decameter scale, which is consistent with the frequent observation of transient features in HF radar backscatter in the cusp ionosphere [*Milan et al.*, 1999]. *Milan et al.* [2005] also showed a close correspondence between amplitude scintillations of 250 MHz satellite beacon signals and SuperDARN backscatter power at 10 MHz. It should be pointed out that 250 MHz satellite beacon signals are more severely affected by scintillation than the 1575.42 MHz (GPS L1) band. We can also see this effect in Figures 5d and 6d, where scintillation is stronger at 1242–1248 MHz (GLONASS L2P) than at 1598–1605 MHz (GLONAS L1CA). It is due to the typical nature of irregularity spectra which show a rapid decay toward shorter wavelengths (higher frequencies), see, e.g., Figures 5e, 6e, and 7e.

The production of irregularities requires energy. PMAFs are believed to be caused by transient reconnection at the dayside magnetopause. Magnetic reconnection transfers flux from the solar wind to the magnetosphere and initiates plasma motion in the polar ionosphere [*Cowley and Lockwood*, 1992; *Lockwood et al.*, 1995]. PMAFs have often been interpreted as ionospheric signatures of FTEs [*Sandholt et al.*, 1990, 1993; *Denig et al.*, 1993; *Milan et al.*, 1999, 2000; *Thorolfsson et al.*, 2000]. At the magnetopause FTEs typically have a scale size of one Earth radius in the boundary normal direction [*Saunders et al.*, 1984]. In the ionosphere the FTE flux tube maps to around 100–200 km along the meridian [*Southwood*, 1985, 1987]. The FTE footprint sets up a PMAF and associated flow shears [*Oksavik et al.*, 2004, 2005], and plasma instabilities continue to structure the plasma down to smaller and smaller spatial scales. The structuring can only continue if energy is input into the system. As we have shown in Figures 3 and 4, the scintillation is co-located with the PMAF, indicating a close relation to their anticipated energy source, transient reconnection at the magnetosphere. A statistical study by *Prikryl et al.* [2015] also suggests that enhanced phase scintillation is highly collocated with regions that are known as ionospheric signatures of the coupling between the solar wind and magnetosphere.

Once formed, the irregularities may cause problems for radio communication and navigation signals, like the severe phase scintillation, loss of signal lock, and cycle slips. It is therefore interesting to note that a process initially starting at the Earth's magnetopause tens of thousands of kilometres away may have impact at much smaller scales in the ionosphere (down to a few hundreds of meters, possibly also smaller) and cause problems of potential importance for society. A particular challenge with PMAF events and their associated plasma irregularities is that they often move with high speeds exceeding 1 km/s. With newly installed multi-constellation receivers offering wide and dense coverage we can now track the scintillation from these disturbances, which would otherwise be smeared out in statistical data sets.

## 5. Concluding Remarks

In this paper we have presented two examples from the cusp ionosphere over Svalbard where bright poleward moving auroral forms (PMAFs) are observed to be associated with severe phase scintillation and strong plasma irregularities at spatial scales of a few hundred meters to a few kilometers. Using a combination of ground-based optical instruments and a newly installed multiconstellation navigation signal receiver we tracked an area of enhanced phase scintillation that was co-located with two PMAFs and moved into the polar cap. Both events affected signals from GPS and GLONASS. One bright PMAF, where the 557.7 nm exceeded the 630.0 nm intensity, also coincided with the steep TEC gradient on the leading edge (poleward side) of a polar cap patch causing a cycle slip; i.e., a change of more than 1.5 TECU per second [*Prikryl et al.*, 2010]. At the same time the receiver also experienced a temporary loss of lock which compromised the GPS L2Y signal (1227.60 MHz) of one spacecraft. The loss of lock appears to have occurred twice in less than one minute, each loss of lock lasting 3–5 s. It shows that PMAF events can cause important space weather effects in the polar ionosphere. Although several polar cap patches were also observed in the TEC data, the scintillation was much stronger from the PMAF events, which also appear to be associated with stronger irregularities than the surrounding cusp ionosphere. It suggests that the structured particle precipitation of a bright PMAF event is an important source for plasma irregularities in the cusp ionosphere, at least at kilometer to hectometer scale [*Oksavik et al.*, 2012].

**Acknowledgments**

The interplanetary magnetic field and solar wind data were provided by the NASA OMNIWeb service (http://omniweb.gsfc.nasa.gov). The UiO ASI data are available at http://tid.uio.no/plasma/aurora. The scintillation data may be obtained from Kjellmar Oksavik (e-mail: kjellmar.oksavik@uib.no). This project has been supported by the Research Council of Norway under contracts 212014, 223252, and 230935.

## Erratum

In the originally published version of this article, the first equation in section 2, "Instrumentation," contained an error. The error has since been corrected, and this version may be considered the authoritative version of record.